\magnification=\magstep1

\font\chhdsize=cmbx12 at 12.4pt
\font\TITLEsize=cmr12 at 18pt
\font\footnotesize=cmr10 at 8pt
\font\abstractsize=cmr10 at 9pt


\def\startbib{\def\biblio{\bigskip\medskip
  \noindent{\chhdsize References}\bgroup\parindent=2em}}
\def\endbib{\edef\biblio{\biblio\egroup}}
\def\BIBlbl#1#2{\global\advance\BIBno by 1
  \edef#1{\number\BIBno}
    \global\edef\biblio{\biblio\medskip\item{[\number\BIBno ]}#2\par}}

 \newcount\BIBno \BIBno=0


 \newcount\equno \equno=0
 \newcount\chno \chno=1

\def\eqlbl#1{\global\advance\equno by 1
  \global\edef#1{{\number\chno.\number\equno}}
  (\number\chno.\number\equno)}


\def\qed{\hfil\hbox to 0pt{}\ \hbox to 2em{\hss}\
         \hbox to 0pt{}\hskip-2em plus 1fill
         \vrule height6pt depth1pt width7pt\par\medskip}

\def\eqed{\hfil\hbox to 0pt{}\ \hbox to 2em{\hss}\
          \hbox to 0pt{}\hskip-2em plus 1fill
          \vbox{\hrule height .25pt depth 0pt width 7pt
            \hbox{\vrule height 6.5pt depth 0pt width .25pt
              \hskip 6.5pt\vrule height 6.5pt depth 0pt width .25pt}
            \hrule height .25pt depth 0pt width 7pt}\par\medskip}


\startbib
\BIBlbl\Bell{
	 J.S. Bell,
		{\it Speakable and unspeakable in quantum mechanics},
	Cambridge University Press, Cambridge, UK (1987).}
\BIBlbl\Goldstein{
        S. Goldstein,
                ``Bohmian mechanics and quantum information,''
        {\it this volume.}}
\BIBlbl\DGZ{
        D. D\"urr, 
	S. Goldstein, 
	and 
	N. Zangh\`{\i}, 
	{\it{Quantum equilibrium and the origin of absolute uncertainty}}, 
	J. Stat. Phys. {\bf{67}},  843-907 (1992).}
\BIBlbl\Feynman{
        R. P. Feynman,
                {\it The Character of Physical Law}, MIT Press (1965).}
\BIBlbl\Bohm{
	D. Bohm, 
		``A suggested interpretation of the quantum
			theory in terms of ``hidden'' variables. Part I,''
	{\it Phys. Rev.} {\bf 85}, 166-179 (1952);
	         ``Part II,''
	{\it ibid.}, 180-193 (1952).}
\BIBlbl\Madelung{
	E. Madelung, 
		``Quantentheorie in hydrodynamischer Form,''
	{\it Z. Phys.} {\bf 40},  322-326 (1926).}
\BIBlbl\Ferris{
        T. Ferris,
        {\it The Whole Shebang: A State-of-the-Universe(s) Report},
        Touchstone, New York (1997).}
\BIBlbl\deBroglieSOLVAY{
        L.V. de Broglie, 
        ``La nouvelle dynamique des quanta,''
        in {\it Cinqui\`eme Conseil de Physique Solvay (Bruxelles 1927)},
        ed. J. Bordet, (Gauthier-Villars, Paris, 1928); 
        English transl.: ``The new dynamics of quanta'', p.374-406 in:
        G. Bacciagaluppi and A. Valentini,
        {\it Quantum Theory at the Crossroads}, (Cambridge Univ. Press, forthcoming).}
\BIBlbl\deBroglieDOUBLEsol{
       L.V. de Broglie, 
	        ``La structure de la mati\`ere et du rayonnement et la m\'ecanique ondulatoire,''
	{\it Comptes Rendus} {\bf 184}, 273-274 (1927);
	        ``La m\'ecanique ondulatoire et la structure atomique de la mati\`ere et du rayonnement,''
        {\it J. Phys. et Rad.} {\bf 8}, 225-241 (1927);
        {\it Une tentative d'interpr\'etation causale et non lin\'eaire de la m\'ecanique ondulatoire: 
                la th\'eorie de la double solution,} Gauthier-Villars, Paris  (1956).}
\BIBlbl\Messiah{
        A. Messiah,
                {\it M\'ecanique quantique}, Tome 1,
        Dunod, Paris (1969).}
\BIBlbl\Hiley{
        B.J. Hiley, 
                ``On the relationship between the Wigner-Moyal and Bohm approaches to
                quantum mechanics: A step to a more general theory?,''
        {\it this volume.}}
\BIBlbl\WignerA{
        E.P. Wigner, 
        ``On the Quantum Correction for Thermodynamic Equilibrium,''
        {\it Phys. Rev.} {\bf 40}, 749-59 (1932).}
\BIBlbl\Moyal{
        J.E. Moyal, 
                ``Quantum Mechanics as a Statistical Theory,''
        {\it Proc. Cam. Phil. Soc.} {\bf 45}, 99-123 (1949).}
\BIBlbl\Born{
	M. Born,
	         ``Zur Quantenmechanik der Stossvorg\"ange,''
	{\it Z. Phys.} {\bf 37},  863-867 (1926);
	         ``Quantenmechanik der Stossvorg\"ange,''
	{\it Z. Phys.} {\bf 38},  803-827 (1926).}
\BIBlbl\WignerB{
        E.P. Wigner, 
        ``Interpretations of quantum mechanics,''
        Lectures given in the
                physics dept. of Princeton University during 1976; revised for publication 1981;
        pp. 260--314 in {\it Quantum Theory and Measurement}, J.A. Wheeler and W.H. Zurek, eds.,
        Princeton Univ. Press, Princeton (1983). (The relevant passages are on p.262 and 290.)}
\BIBlbl\BohmII{
	D. Bohm, 
		``Reply to a criticism of a causal re-interpretation of
                the quantum theory,''
	{\it{Phys. Rev.}} {\bf{87}}, 389-390 (1952);
        ``Comments on an Article of Takabayashi concerning the 
          formulation of quantum mechanics with classical pictures,''
         {\it{Prog. Theor. Phys.}} {\bf{9}}, 273-287 (1953).}
\BIBlbl\BST{
	D. Bohm, 
	R. Schiller,
	and
	J. Tiomno,
        ``A causal interpretation of the Pauli equation (A),''
	{\it{Nuovo Cim. Suppl.}} {\bf{1}}, 48-66 (1955).}
\BIBlbl\BohmBub{
        D. Bohm,
        and
        J. Bub,
                ``A proposed solution of the measurement problem in quantum mechanics
                by a hidden variable theory,''
        {\it{Rev. Mod. Phys.}} {\bf{38}}, 453-469 (1966).}
\BIBlbl\BohmHileyBOOK{
	D.~Bohm,
	and
	B.J.~Hiley,
		{\it{The undivided universe},}
	Routlege, London, UK (1993).}
\BIBlbl\Kie{
	M.K.-H. Kiessling, 
		``Electromagnetic field theory without divergence problems.
		        1. The Born legacy,''
	{\it J. Stat. Phys.} {\bf 116}, 1057-1122 (2004);
		        ``Ditto. 2. A least invasively quantized theory,''
	{\it J. Stat. Phys.} {\bf 116}, 1123-1159 (2004).}
\BIBlbl\BoIn{
	M. Born 
	and
	L. Infeld,
        	``{Foundation of the new field theory},''
	{\it Nature} {\bf{132}}, 1004 (1933);
        ``Ditto,''
	{\it{Proc. Roy. Soc. London}} {\bf{A 144}}, 425-451 (1934).}
\BIBlbl\berndlETal{
	K.~Berndl, 
        D.~D\"urr, 
	S.~Goldstein, 
	and 
	N.~Zangh\`{\i}, 
        ``{Nonlocality, Lorentz invariance, and Bohmian quantum theory,}''
	{\it{Phys. Rev. A}} {\bf{53}}, 2062-2073 (1996).}
\BIBlbl\munchberndlETal{
        D.~D\"urr, 
	S.~Goldstein, 
	K.~M\"unch-Berndl, 
	and 
	N.~Zangh\`{\i}, 
        ``{Hypersurface Bohm-Dirac models,}''
        {\it Phys. Rev. A} {\bf 60}, 2729-2736 (1999).}
\BIBlbl\Tumulka{
        R.~Tumulka,
                ``The ``Unromantic Pictures'' of Quantum Theory,'' 
        {\it{J. Phys. A}} {\bf{40}}, 3245-3273 (2007).}
\endbib


\centerline{ \TITLEsize  Misleading signposts along the}
\centerline{{\TITLEsize  de Broglie-Bohm road to quantum mechanics}\footnote{$^{^*}$}{\footnotesize
  Version of Sept.19, 2007.\              Dedicated to Jeffrey Bub on occasion of his 65th birthday.
$\phantom{--------}$
\copyright (2007) The author. Reproduction of this article, in its entirety, 
is permitted for noncommercial purposes.}} 
\medskip
		
\centerline{\it Michael K.-H. Kiessling}
\smallskip
\centerline{Department of Mathematics}
\centerline{Rutgers, The State University of New Jersey}
\centerline{110 Frelinghuysen Rd., Piscataway, NJ 08854}
\centerline{email: miki@math.rutgers.edu}
\smallskip

\centerline{\bf Abstract} \noindent
{\abstractsize
Eighty years after de Broglie's, and a little more than half a century after Bohm's 
seminal papers, the de Broglie--Bohm theory (a.k.a. Bohmian mechanics), which is presumably 
the simplest theory which explains the orthodox quantum mechanics formalism, has 
reached an exemplary state of conceptual clarity and mathematical integrity.
        No other theory of quantum mechanics comes even close. 
        Yet anyone curious enough to walk this road to quantum mechanics is soon being confused 
by many misleading signposts that have been put up, and not just by its detractors, but unfortunately 
enough also by some of its proponents.
        This paper outlines a road map to help navigate ones way.}
\bigskip

\noindent{\chhdsize 1\quad De Broglie-Bohm theory in a nutshell}
\smallskip

        Already in its basic version without spin and magnetic interactions the theory yields a quite 
faithful Galilei covariant realist caricature of our world.
	Thus, there is a three-dimensional space, represented by ${\bf R}^3$, and there is time, 
represented by ${\bf R}$; at any instant $t$ in time this space is populated by
a fixed number $N\gg 1$ of point particles at locations $Q_k(t)\in{\bf R}^3$, $k=1,...,N$; as time goes 
on, these particles move according to the law of motion
$$
{{\rm d} \over {\rm d}t} Q_k(t) = v_k^\psi(Q(t),t),
\eqno\eqlbl\guidingEQ
$$
where $v^\psi(\,\cdot\,,t):{\bf R}^{3N}\to{\bf R}^{3N}$ is a velocity field at time $t$ on the space of 
generic configurations $q\in{\bf R}^{3N}$, the $k$-th ${\bf R}^3$ component of which reads
$$
v_k^\psi(q,t) = {\rm Re} 
\Big((\psi^\dagger\psi)^{-1} \psi^\dagger\big(-i{\textstyle{\hbar\over m_k}}\nabla_k\big) \psi\Big) (q,t);
\eqno\eqlbl\vpsi
$$
the complex-valued field $\psi(\,\cdot\,,t): {\bf R}^{3N}\to {\bf C}$ is Schr\"odinger's wave function 
at time $t$ for the $N$-particle system (a.k.a. its de Broglie wave), 
satisfying Schr\"odinger's wave equation
$$
i \hbar {\partial_t \psi(q,t)} = H \psi(q,t).
\eqno\eqlbl\waveEQ
$$
	A quite faithful model of our world obtains when as Hamiltonian we take
$$
H = \sum_{1\leq k\leq N} {{1\over 2m_k}} (-i\hbar\nabla_k)^2+
\sum\sum_{\hskip-.7true cm 1\leq k< l \leq N} {e_k e_l - Gm_km_l \over |q_k-q_l|} ,
\eqno\eqlbl\H
$$
with $e_k =-e$ and $m_k = m$ for $k=1,...,N_e$, while $e_k = Z_ke$ with $Z_k$ a natural 
number and $m_k \gg m$ for $k=N_e+1,...,N$, and we also demand $\sum_{k=1}^N e_k =0$; otherwise
the symbols have their standard meaning.
\vfill\eject

	The dynamical equations (\guidingEQ), (\waveEQ) pose an initial value problem and as such 
need to be supplemented by initial data, say at $t=0$,  viz. by  $\psi(\,\cdot\,,0)$ and $Q(0)$.
	For $\psi(\,\cdot\,,0)$ we demand it be antisymmetric w.r.t. permutations of particle indices
in $\{1,...,N_e\}$ (the electrons), while for other groups of particle indices (the various positive
nucleus species) one imposes symmetry or antisymmetry w.r.t. permutations (we don't need to bother 
to be more specific here).
	Also, the initial $\psi$ has to be sufficiently regular (formally a $C^\infty$ vector for $H$).
	For $Q(0)$ we demand that it be {\it typical} with respect to $|\psi(\,\cdot\,,0)|^2 d^{3N}q$.

        The model can easily be generalized to account for the effects of so-called 
``external''\footnote{$^1$}{\footnotesize
                In a model of a universe, ``external'' is an odd concept. 
                Technically, it means that these electromagnetic fields are simply given in 
                addition to the Coulomb pair interaction in $\scriptstyle H$.}
electromagnetic fields, as well as for electron spin, through the replacements 
$-i\hbar \nabla_k \to \sigma_k\cdot(-i\hbar\nabla_k - e_k A(q_k,t)/c)$
and $i\hbar \partial_t \to i\hbar\partial_t - e_k \phi(q_k,t)$, and with 
$\psi(\,\cdot\,,t)\in {\bf C}^{{\bf R}^{3N}\times \{-1,1\}^{N_e}}$ now an $N_e$-particle Pauli spinor
w.r.t. the electron indices, such that $\psi^\dagger\psi$ now is the inner product over spinor space 
degrees of freedom, and $\sigma_k$ the 3-vector of Pauli matrices 
acting on the $k$-th electron index (similar modifications can be arranged for the nuclei).

        As John Bell [\Bell] put it, to the extent that (non-relativistic) orthodox quantum mechanics 
makes unambiguous predictions, the de Broglie-Bohm  model makes the same predictions.
        What's more, it provides a theory of our world: a three-dimensional world populated by point 
objects (particles) which move according to a (non-Newtonian) law of motion, their motion being such 
as to produce the overall material structures and processes we happen to recognize in the world we 
live in.
	Paraphrasing Dirac, the de Broglie-Bohm model accounts for all of chemistry and most of 
physics, from the existence of atoms and their main energy spectra (not the emission and absorption 
of (the relativistic) photons, though) to the existence and motion of things like planets and their moons
---
though Dirac of course said something like this about orthodox nonrelativistic quantum mechanics.
	Exactly how this theory accounts for all that, and without any magic, is now a 
fascinating matter of logical deduction, of which we will give no details here.
	You can read more about it in the exposition by Sheldon Goldstein in this volume 
[\Goldstein], and in [\DGZ].
\medskip

\noindent{\chhdsize 2\quad It's the ontology, stupid!}\footnote{$^\dagger$}{\footnotesize The title is
                                obviously inspired by a famous Clinton campain slogan from the nineties
                                and should not be misconstrued as a personal attack by me on anyone.}
\chno=2
\smallskip

        The purpose of the title of this section is to focus the attention on what is the truly 
important distinction of the de Broglie-Bohm theory, namely that {\it this quantum theory is about 
something objectively real}, in this case point particles which move according to a new law of motion, 
(\guidingEQ) with r.h.s. given by (\vpsi), which complements Schr\"odinger's wave equation (\waveEQ). 
        It is this aspect, and this aspect alone, which restores intelligibility to the whole affair of 
nonrelativistic quantum physics, of which Feynman [\Feynman] famously said:
``I think I can safely say that nobody today understands [orthodox]  quantum mechanics.'' 
        While all of orthodox quantum physics suffers from a `reality-deficiency-disorder' known as 
``the measurement problem,'' whose symptoms are ameliorated by magical spells (`whenever one 
performs a measurement, then (a miracle occurs and) reality becomes manifest, and only then'), the 
de Broglie-Bohm theory by contrast is a healthy realist theory of a nonrelativistic objective universe.
        This asset of the de Broglie-Bohm theory cannot be treasured too much.

        Curiously enough, though, the theory is generally not well appreciated.
        The following quote, taken from the Encyclopedia Britannica (quoted in [\Goldstein]), 
is typical in this regard:
\smallskip

        ``Attempts have been made by Broglie, David Bohm, and others to construct theories based 
on hidden variables, but the theories are very complicated and contrived. 
        For example, the electron would definitely have to go through only one slit in the two-slit
experiment. 
        To explain that interference occurs only when the other slit is open, it is necessary to 
postulate a special force on the electron which exists only when that slit is open. 
        Such artificial additions make hidden variable theories unattractive, and
there is little support for them among physicists.''
\smallskip

        This quotation from the Encyclopedia Britannica is quite illuminating in (at least) two 
ways.

        First, the author of these Encyclopedia Britannica lines, and the editor who approved 
them,\footnote{$^2$}{\footnotesize
                The Encyclopedia Britannica section on quantum physics is presided over
                by a Nobel laureate in physics as special editor, and its entries are written by 
                professional physicists of some standing (just so you know I am not unfairly criticizing
                the production of some poorly trained and overtaxed professional staff writer).}
apparently recognize that in the de Broglie-Bohm theory an electron actually does have a position 
(and, therefore, definitely does ``go through only one slit in the two-slit experiment''),  but 
it is also quite apparent that it is either not recognized that this endows the theory with an ontology,
something that is missing from the orthodox theory, or if it is recognized, then it is obviously not 
treasured at all!
        I can only hope that what's expressed in the title of this section will eventually sink in.

        Second, that author, and the editor, must have gotten their `insights' about the de Broglie-Bohm 
theory from some really ``very complicated and contrived'' (looking) exposition, for he or she could 
not possibly be writing such things about equations (\guidingEQ)--(\H).
        It sounds like there is a lesson to be learned here!
\medskip

\noindent{\chhdsize 3\quad Newton's ghost}
\chno=3
\smallskip
        
        Let's be clear about this: the Encyclopedia Britannica entry about hidden variable theories
totally misrepresents the de Broglie-Bohm theory --- except for the claim that an electron 
has to ``go through only one slit in the two-slit experiment,'' and that 
``there is little support for them [hidden variable theories] among physicists.''
        It misrepresents the theory on three counts:

$\bullet$ The model (\guidingEQ)--(\H) is {\it manifestly not} ``very complicated,'' 
certainly not by absolute standards, and especially not when compared with the orthodox formalism 
of nonrelativistic quantum mechanics, which shares with it the Schr\"odinger equation (\waveEQ) 
with $H$ given by (\H), but which lacks (\guidingEQ) and instead supplies a literally open-ended 
list of so-called measurement postulates --- by that measure the de Broglie-Bohm model is even 
infinitely more simple. 

$\bullet$ Neither is the model (\guidingEQ)--(\H) ``contrived,'' for not only is the Schr\"odinger equation 
(\waveEQ) with $H$ given by (\H) the same as in the orthodox theory, but also the velocity field (\vpsi) 
is simply what in the orthodox theory is called the probability current density vector $j(q,t)$ divided
by the probability density $\rho(q,t) = |\psi|^2(q,t)$, so $v^\psi = j/\rho$.
	Given these established mathematical ingredients of the orthodox theory, if one now tries 
to make sense of the {\it orthodox talk} about particles, then it isn't exactly ``contrived'' 
to contemplate that particles actually do have a position, so that $N$ of them correspond to an actual 
point $Q(t)$ in configuration space --- and isn't that, anyway, already suggested by the simple fact 
that Schr\"odinger's wave function is a function on $N$-point configuration space?
	Once one has reached this insight, the next logical step is to look for a natural law of motion
for these point particles, and having a velocity field on configuration space essentially `for free' 
already, how ``contrived'' is it to contemplate that this velocity field evaluated at the actual 
point $Q(t)$ {\it actually is} the velocity of $Q(t)$? 
        Well, that's exactly what (\guidingEQ) states. 

$\bullet$ Neither does one have to ``postulate a special force on the electron which exists only when 
that [other] slit is open [too]'' in order ``to explain that interference occurs.''
        All that is postulated beyond  Schr\"odinger's equation (\waveEQ), is that 
the electron is guided by the velocity field (\vpsi) obtained from the solution
$\psi(q,t)$ of Schr\"odinger's equation (\waveEQ). 
        Moreover, $\psi$ develops an interference pattern only when both slits are open, 
a fact that is inevitably reflected in the manner how $\psi$ guides the electron.

        Needless to say, by misrepresenting the de Broglie-Bohm theory one also
misleads uninitiated readers: if I hadn't known anything about that theory
and read about it for the first time in the authoritative Encyclopedia Britannia, I would surely
have gotten the impression that the theory {\it is} contrived and --- presumably --- very complicated.
        Incidentally, something like that actually did happen to me, though it was my professor in the 
course ``Philosophical problems of physics'' which I attended as a student, who told me
basically the same things you read in the Encyclopedia Britannica article, and it did turn me against
this theory.\footnote{$^3$}{\footnotesize 
                        Lest the reader gets the impression that this course was a waste of
                        my time, I hasten to add that it was one of the more important courses 
                        about quantum physics that I attended. Prof. B\"uchel was reassuringly
                        dismissive of the Copenhagen interpretation and very critical of the prevailing
                        logical-positivistic attitude in physics. Yet also he had the wrong impression
                        of what de Broglie-Bohm theory is.}
        That was before I took a closer look myself. 
        Bell's book [\Bell] was a revelation, and the writings by D\"urr, Goldstein, Zangh\`{\i} 
and their collaborators clarified matters for me considerably.

        But who is responsible for such misleading misrepresentations of the theory? 
        Of course, the ultimate responsibility always lies with their author, but I think it is 
fair to say that blame also should go to some of the main actors and their supporting cast for
creating (unintentionally, I suppose) certain {\it{misperceptions}} of the theory. 
        Thus, the original source of the particular
misperception that the theory is postulating ``special forces,'' a Newtonian concept, hence a misguided
attempt to ``return to the womb of classical physics,''\footnote{$^4$}{\footnotesize
                ``It's haughty of me, but I do think that myopia is running rampant in our community 
                ... not only in the reading of our little article,
                but in the much bigger picture---our hopes for a physical theory.
                Bohmism is an example of such a dull point of view:  If we can just
                return to the womb of classical physics, everything will be oh so
                much more warm and comfortable.  Yuck!'' (From a letter by C.A. Fuchs to M.B. Ruskai
                on 24 March 2001; arXiv:quant-ph/0105039.)}
alas, is [\Bohm]. 
        Indeed, while Bohm [\Bohm] does write down the theory which I've presented in section 1 --- 
he does so in a small paragraph on p.6 of Part I, and again in the first paragraph of Part II,
of his two-part article --- much more prominence, by far, is given to a derived concept, which 
is used as a point of departure for a speculative generalization of the theory.
        Namely, by taking the time derivative of (\guidingEQ), then using (\vpsi), (\waveEQ), 
(\H),\footnote{$^5$}{\footnotesize
                To be more accurate, Bohm does not explicitly give the interaction term in (\H) but
                just uses the more general $\scriptstyle V(q_1,q_2,...)$ (in our notation),
                but this is irrelevant to our argument.}
a Newtonian-type second order equation of motion is obtained which exhibits, besides the familiar 
Coulomb and Newton forces, the gradient of a term that depends on $|\psi|$ and its second spatial 
derivatives.
        This term, which was already well-known (see, e.g., [\Madelung]) through 
rewritings of Schr\"odinger's equation (\waveEQ) into a system of equations, using the polar 
decomposition of $\psi = |\psi|e^{i\Phi}$ (see next section), Bohm called the ``quantum potential.'' 
        Of course, in order that {\it everything else remains equal} (i.e. the predictions derived
from it remain unaltered), the first order equation (\guidingEQ) evaluated at the initial time still 
has to be imposed as an initial condition, and then it propagates for all times.
        So, with (\guidingEQ) understood as valid initially, Bohm's reformulation of (\guidingEQ)--(\H) 
into a Newtonian equation of motion coupled to Schr\"odinger's equation is strictly equivalent to 
(\guidingEQ)--(\H). 
        But as such it is unnecessarily more complicated, for $\psi$ cannot be eliminated by taking 
the extra time derivative of (\guidingEQ).\footnote{$^6$}{\footnotesize
		By comparison, Newtonian point mechanics in the 
		Hamilton-Jacobi formalism operates with the same guiding equation 
		(\guidingEQ) (when written in polar decomposion), but with the Hamilton-Jacobi 
		partial differential equation for $\scriptstyle \Phi_{HJ}$ in place of Schr\"odinger's 
                equation (\waveEQ); while $\scriptstyle \psi$ cannot be eliminated from the equations 
                of motion for $\scriptstyle Q$ by differentiation of (\guidingEQ),  the Hamilton-Jacobi 
                phase field $\scriptstyle \Phi_{HJ}$ on configuration space can.}
        In any event, there is no doubt in my mind that Bohm's pseudo-Newtonian reformulation has 
given more casual readers the impression that he tries to derive quantum mechanics from Newtonian 
point mechanics by postulating new special forces derived from a new potential, the quantum potential.
        Here is another victim (who gets it even more upside down, but who in his book
actually seems quite sympathetic to Bohm) [\Ferris]:
	``Bohm was obliged to invent an agency -- a guiding wave -- to manipulate the 
particles. He called this guiding wave the ``quantum potential''.'' 

        Bohm's apparent motivation for presenting matters his way, as I understand [\Bohm], is his hope
that by dropping `(\guidingEQ)-evaluated-at-the-initial-time' as initial condition the so-obtained 
freedom to choose initial conditions also for the velocities will yield new physics in form of a
{\it generalization of quantum mechanics}.
        This is certainly a legitimate hope to entertain,  but what a 
considerable {\it lapse of judgement} it was, indeed, to pursue such a speculative idea which says 
that {\it not everything else is equal} in a two-part paper whose title declares that its main purpose 
is to suggest an ``interpretation of {\it the quantum theory} in terms of ``hidden'' 
variables''\footnote{$^7$}{\footnotesize
        Incidentally, ``interpretation of the quantum theory'' is also not a lucky choice of words,
        for Bohm does not just interpret the theory in a new way, he presents a deeper theory which
        explains (nonrelativistic) quantum theory, namely the very theory de Broglie was pursuing
        in the 1920s before joining the Copenhagen bandwagon as a fallout from the 5th Solvay
        conference in 1927.}
(emphasis mine) --- and which actually achieves the declared goal, for the first time ever!
        I wish Bohm had written his two-part paper [\Bohm] entirely in terms of equations
(\guidingEQ)--(\H) and relegated all the speculation about a possible generalization of
quantum mechanics based on the second order Newtonian equation with its ``quantum potential''
to a separate, third paper entitled ``A speculative generalization of quantum theory inspired 
by my [Bohm's] previous two papers.'' 
        Of course, the reality is what it is, but next time you contemplate emphasizing 
`the importance of the quantum potential in order to understand quantum mechanics,' 
think again!

        I am not sure how much the original main actor, de Broglie, who in [\deBroglieSOLVAY] 
states the de Broglie-Bohm $N$-particle model for the first time, has contributed to the 
misconception that the model is Newtonian through his later comments on Bohm's papers.
        Yet he certainly contributed his own share to the general confusion about the model by 
not working things out properly, and by abandoning the pursuit of his model right after the 5th Solvay 
conference,\footnote{$^8$}{\footnotesize 
                In his acceptance speech for the Nobel prize in physics 2 years later, de Broglie 
                not only omits mentioning his original ideas that particles are guided by a pilot
                wave, he actually presents the Copenhagen doctrine that particles do not 
                have a position at all times.}
resuming it only after Bohm's 1952 papers, and then not in a consequential way but instead by
chasing after his own favorite speculative ideas about the model's origins in his
``theory of the double solution'' [\deBroglieDOUBLEsol].
        But that's another story about which I say a few things in the last section.
\medskip

\noindent{\chhdsize 4\quad Polar disorder}
\chno=4
\equno=0
\smallskip
        
        Both de Broglie [\deBroglieSOLVAY] and Bohm [\Bohm] chose a polar representation of the 
Schr\"odinger wave function, viz. $\psi =  |\psi|e^{i\Phi}$ (though de Broglie frequently
writes only the real part of it; why, I don't know), through which the guiding equation (\guidingEQ)
takes the particularly simple form
$$
{{\rm d} \over {\rm d}t} Q_k(t) = {\hbar\over m_k} \nabla_k\Phi(q,t)\Big|_{q=Q(t)}
\eqno\eqlbl\guidingEQpolar
$$
while Schr\"odinger's equation (\waveEQ) with (\H) becomes a coupled system for $|\psi|$ and $\Phi$:
the continuity equation
$$
\partial_t |\psi|^2(q,t) = -  \sum_{1\leq k\leq N}
\nabla_k\cdot\left( |\psi|^2(q,t) {\textstyle{\hbar\over m_k}}\nabla_k\Phi(q,t)\right) 
\eqno\eqlbl\cEQ
$$
and a generalization of the classical Hamilton-Jacobi equation, viz.
$$
\hbar\partial_t \Phi(q,t) = - 
\sum_{1\leq k\leq N} 
{{\hbar^2\over 2m_k}} \left(|\nabla_k\Phi(q,t)|^2 - {\Delta_k|\psi|\over |\psi|}(q,t)\right)
- \sum\sum_{\hskip-.7true cm 1\leq k< l \leq N} {e_k e_l - Gm_km_l \over |q_k-q_l|} .
\eqno\eqlbl\HJqEQ
$$
        Without the $\Delta|\psi|/|\psi|$ term (\HJqEQ) would be the Hamilton-Jacobi equation for
$\Phi_{HJ}$; the $\Delta|\psi|/|\psi|$ term is what Bohm called the ``quantum potential.''
        Save some technical issues regarding the question 
whether globally differentiable functions $|\psi|$ and $\Phi$ can be found for a differentiable 
$\psi$, at least {\it locally} Schr\"odinger's equation (\waveEQ) with (\H) is entirely 
equivalent to (\cEQ) and (\HJqEQ). 
        It should be noted that the gain in simplicity of the guiding equation is entirely offset
by the gain in complexity and nonlinearity when rewriting (\waveEQ) with (\H) into (\cEQ) and
(\HJqEQ). 
        Rewriting (\guidingEQ)--(\H) into the polar form (\guidingEQpolar)--(\HJqEQ) is 
very helpful when studying the classical limit of the de Broglie-Bohm model, but otherwise it makes 
things unnecessarily complicated.
        And, of course, if one wishes to compare the de Broglie-Bohm theory with orthodox
quantum mechanics, the fact that the orthodox equation (\waveEQ), with (\H), is perfectly
fine and doesn't need to be transformed in any way, is quite important.
        
        Apropos the classical limit: in [\Messiah] Messiah uses  the polar representation of 
Schr\"oding\-er's equation and notes that whenever the $\Delta|\psi|/|\psi|$ term is 
negligible in (\HJqEQ) then the classical Hamilton-Jacobi equation results, decoupled from the 
continuity equation.
        Strangely enough, Messiah then claims that thereby one has obtained classical Newtonian point 
mechanics --- but this is of course a {\it non sequitur} without the guiding equation!
        The classical Hamilton-Jacobi PDE without the guiding equation (\guidingEQpolar) is {\it not}
equivalent to Newtonian point mechanics, because this PDE, alone or with the continuity equation, 
does not supply the ontology of Newton's theory: point particles whose velocities are determined by the 
configurational velocity field components $m_k^{-1}\nabla_k\Phi$ {\it{evaluated at the actual configuration}}.

        In the same vein, if in a paper on quantum mechanics the polar representation of the 
Schr\"odinger equation is used, either in disguise or (semi-)explicitly, and the `configurational
velocity field' $v=j/\rho$ written as $v_k=m_k^{-1}\nabla_k\Phi$, yet that paper 
does not contain the particle ontology {\it together} with the guiding equation (\guidingEQpolar) 
in some form (so that the `configurational velocity field' $v$ is not used to define the motion of
an actual configuration point),  then it is a {\it non sequitur} to assert that such a paper contains
the essential equations of Bohm's papers [\Bohm].
        Such a {\it non sequitur} can actually be found in this volume in the 
paper by Hiley [\Hiley].
        Hiley discusses the {\it phase space representation} of the Schr\"odinger equation,
which goes back to the works of Wigner [\WignerA] and Moyal [\Moyal], though they pursued
different goals.
        In this approach one works with the Wigner transform of $\psi$, 
$$
F(p,q,t) = {1\over h^{3N}}\int \psi^*(q  - {\textstyle {1\over 2}}q^\prime,t)
                                \psi(q + {\textstyle {1\over 2}}q^\prime,t)
                                e^{ip\cdot q^\prime/\hbar}{\rm d}^{3N}q^\prime.
\eqno\eqlbl\WignerF
$$
        Since the time evolution for $F(p,q,t)$ is equivalent to Schr\"odinger's equation for $\psi(q,t)$, 
it should come as no surprise that if one manipulates it long enough, as did Moyal and as does Hiley, then 
the polar (\cEQ), (\HJqEQ) show up again. 
        Moreover, Moyal's paper also contains the `configurational velocity field' $v=m^{-1}\nabla\Phi$, and
he wrote his paper truly in the spirit of searching for a law of motion 
for actual particles, though his hope was that the quasi-Markovian equation on phase space for 
the Wigner-Moyal function (\WignerF) would reveal a stochastic process on phase space that 
describes the motion of the particles. 
        In this sense Moyal had his heart in the right place.
        His hopes were dashed by the fact, unbeknownst to him at the time it seems, that the Wigner-Moyal
function (\WignerF) is not always positive and as such cannot be interpreted as an ensemble 
probability density function, and the quasi-Markovian equation for it not as a Kolmogorov equation.
        However, Moyal did not have the guiding equation (\guidingEQpolar) in his paper!
        Having a particle ontology {\it in mind}\footnote{$^9$}{
                        \footnotesize Paraphrasing John Stewart Bell [\Bell]:
                                `It has to be in the mathematics, not just in the talk!' 
                        Indeed, if having a particle ontology in mind were enough, then we
                        would have to call it the `de Broglie-Bohm-Born-Moyal-Wigner-...'
                        model, instead of de Broglie-Bohm, for also Born in [\Born] and
                        Wigner in [\WignerB] talk very explicitly about $\scriptstyle \psi$ being a 
                        ``F\"uhrungsfeld'' which is guiding the particles.
                        All the same, both Born and Wigner are critical of attempts to supply an 
                        actual guiding equation, though for different reasons: Born gives 
                        logical-positivistic philosophical 
                        reasons, while Wigner argues conceptually: since $\scriptstyle |\psi|^2$, 
                        which is an ensemble probability density (viz. standing for our ignorance), 
                        enters the ``quantum'' Hamilton-Jacobi equation (\HJqEQ) for $\scriptstyle \Phi$, 
                        it follows that $\scriptstyle \nabla\Phi$ cannot possibly guide an 
                        $\scriptstyle{individual\ member}$ of the ensemble. 
                        But of course $\scriptstyle |\psi|^2$ is $\scriptstyle{not\ fundamentally}$ 
                        a probability density. Now that sounds paradoxical, for Born's law  says that 
                        $\scriptstyle |\psi|^2$ is a probability density. 
                        How the paradox is resolved by the de Broglie-Bohm theory 
                        you  can read in  [\Goldstein] and especially [\DGZ].}
 doesn't mean it's in the theory.
        Wigner in [\WignerA], on the other hand, was not at all looking for any ontology
but instead was concerned with purely practical matters of computing quantum corrections to the classical 
Boltzmann-Gibbs statistical ensemble predictions.

        While it is thus misleading to claim the basic equations of the de Broglie-Bohm model 
were contained in a paper featuring (\cEQ), (\HJqEQ) and $v_k = m_k^{-1}\nabla_k\Phi$ 
{\it without} an ontology, even if you see (\cEQ) and (\HJqEQ) together with $v_k = m_k^{-1}\nabla_k\Phi$ 
in a paper {\it with} an ontology, that doesn't mean you have the de Broglie-Bohm model in front of you.
        As noted earlier, the rewriting of (\waveEQ) with (\H) into polar form (\cEQ) and
(\HJqEQ) was there almost from the beginning.
        Beside de Broglie, also Madelung [\Madelung] made use of it.
        Madelung also had an ontology!
        However, Madelung pointed out that the polar representation gives a system of PDEs which, 
when $N=1$, can be given a {\it fluid-dynamical interpretation}.
        This is very much in
the spirit of Schr\"odinger's concept of $\psi$ as {\it matter waves}. 
        There are no fundamental particles in either Madelung's or Schr\"odinger's interpretation.
        But Madelung's, like Schr\"odinger's, ideas of a physical material continuum ontology 
work only on three-dimensional space (conceptually, that is; they do not produce the correct physics),
while they are {\it absurd} for an equation on $N$-particle configuration space. 
        Oddly enough, while Madelung noted that no such fluid ontology interpretation seems 
feasible when $N>1$,  his model enjoys a large following these days in the semi-conductor 
modeling community. 
\medskip

\noindent{\chhdsize 5\quad Quo vadis?}
\chno=5
\smallskip
\hskip1truecm {\it  Richtiges Auffassen einer Sache und Mi\ss verstehen der gleichen Sache

\hskip1truecm       schlie\ss en einander nicht vollst\"andig  aus. --- Franz Kafka}
\smallskip

        In the previous four sections, I have first presented the de Broglie-Bohm model
(a.k.a. Bohmian mechanics), which is found both in [\deBroglieSOLVAY] and in [\Bohm]
except for the notation and the different representation of Schr\"odinger's equation (cf. my section 4). 
        Next I've emphasized that the model is presumably the simplest and conceptually 
most natural realist completion of non-relativistic orthodox QM, which removes all the paradoxes of
the orthodox theory, and that as such it is a milestone. 
        I have then addressed some misconceptions of the theory, explained what's wrong with them, and 
argued that these misconceptions were (and are) at least facilitated by confusing presentations of 
the theory, and not just by its detractors but, alas,  by some of its very proponents. 
        The lesson one learns from this is how immensely important it is to clearly focus on what 
the theory is about --- its ontology, in this case point particles which move according to a simple 
and obvious (given Schr\"odinger's equation) non-Newtonian law of motion --- and to present the theory 
in a technically clean and simplest possible manner. 
        Ignore this and you confuse others and, perhaps, yourself.
        
        In this last section I want to ask: ``Where does one go from here?''
        After all, the model (\guidingEQ)--(\H) presented in section 1 is non-relativistic, so it
can only be a stepping stone en route to a deeper, more general theory. 
        De Broglie and Bohm were of course aware of this fact.
        Both strived, each one in his own way, to find such a deeper theory. 
        Neither of them held monolithic views about it.

        Bohm, who started out a `Copenhagenist' before his interaction with Einstein eventually turned him
against the Copenhagen doctrine, looked for a deeper theory already in his 1952 papers [\Bohm], where
the de Broglie-Bohm theory of quantum mechanics is presented only as a special case of his unhappy
speculative generalization of it, as I've explained in section 3; yet this work also contains
a generalization of de Broglie-Bohm theory to include photons, in an appendix.
        Shortly after he also knew how to extend the de Broglie-Bohm model to the Dirac equation 
[\BohmII].
        After those initial steps toward a relativistic generalization of the de Broglie-Bohm model
came a longer intermission during which Bohm explored other ontologies.
        For instance, together with his Brazilian collaborators he worked on the extension of the 
Madelung fluid ontology to include spin [\BST]; this paper gives you the impression that Bohm had 
already abandoned the point particle ontology around 1955.
        Then there is his joint work with de Broglie's student Vigier, and his work with Bub [\BohmBub].
        But eventually  Bohm picked up the de Broglie-Bohm model 
again in his collaboration with Hiley [\BohmHileyBOOK].
        Bohm's pet idea became the ``implicate order,'' something I haven't digested. 
        
        De Broglie's path was not straight either, but of course different in many ways.
        He began his career a realist, developing his pet idea of the ``double solution theory'' 
early on, dropping it, then picking it up again after a 25 year intermission [\deBroglieDOUBLEsol].
        In those intermediate 25 years he was a faithful follower of the Copenhagen doctrine;
cf. footnote~8.
        De Broglie's double solution theory itself evolved over time: early on it involved
looking for singular solutions of a system of $N$ one-body Schr\"odinger (or Klein-Gordon) 
equations, suitably constrained, the singularities of which he hoped to show would move according 
to the guiding field obtained from the gradient of the phase of a singularity-free Schr\"odinger 
wave function which solves an associated  Schr\"odinger $N$-body equation; later it involved 
also the Dirac equation and finally nonlinear Schr\"odinger-type equations [\deBroglieDOUBLEsol].
        As is well-known, the study of nonlinear Schr\"odinger (and Klein-Gordon, and Dirac) equations 
on physical spacetime has become quite popular in recent decades, though for different reasons.

        De Broglie's pursuit of the double solution is, in my opinion, another example of a serious
{\it lapse of judgement}, though a quite different one from Bohm's (cf. section 3).
        Namely, it is a great mystery to me why de Broglie 
--- who was thinking about point electrons as singularities in a field on physical space --- 
was thinking of singularities in the field solutions to some Schr\"odinger-type equation. 
        After all, at least in classical electrodynamics point electrons {\it are} already given 
to us as point singularities --- in the classical electromagnetic Maxwell fields; furthermore, 
the energy of an electrostatic field with $N$ of these point singularities (with the infinite 
self-energies subtracted) is precisely the Coulombic interaction term in the Hamiltonian (\H) 
which enters Schr\"odinger's equation (\waveEQ). 

        Incidentally, I have reached a point where it is not inappropriate to say a few words 
about where I myself have been going. 
        The so-called UV problems caused by such point singularities in the electromagnetic fields 
of Maxwell's linear field theory are precisely the point of departure for my own pursuit of an 
intelligible realistic quantum field theory [\Kie].
        Nonlinear  field equations are an important ingredient in the model I am working
with, though not a nonlinear modification of some Schr\"odinger equation but the nonlinear 
Maxwell-Born-Infeld field equations of classical electromagnetic theory [\BoIn].
        These nonlinear field equations remove the infinite self-energies of point electrons
without any artificial cutoffs; thus, point electrons now appear as much milder point defects 
of the electromagnetic field. 
        But the Maxwell-Born-Infeld field equations do not provide the law of motion of those point
defects --- it is here where Hamilton-Jacobi and de Broglie-Bohm formalisms appear to be `just what
the doctor ordered.'
        The works [\Kie] are about the classical and quantum electromagnetic theory of point electrons 
without spin.
        Spin, handled by Dirac's equation, and curved spacetimes have by now been included; this is 
being written up for publication.
        Photons are still somewhat elusive, but I will have something to say about those also.
        This theory is as relativistic as possible, but there is a price to be paid. 

        Namely, I haven't yet mentioned the most striking problem.
        De Broglie-Bohm theory is manifestly non-local, in a manner that 
`infinitely' trumps the extent of non-locality in Newton's point mechanics, 
where the electrical and gravitational forces between different constituents 
decay reciprocally with the square of their separation. 
        Bell's work [\Bell] brought to the fore that the measurement formalism of orthodox 
quantum theory makes the very quantum theory non-local. 
        So the real question, as Bell [\Bell] has emphatically stressed, is how to
replace the ontology-free non-local measurement formalism with an intelligible, realistic, 
physically objective non-local mathematical theory which obeys special, and eventually
general relativity --- the paradigms of locality!

        One intriguing insight which the pursuit of a relativistic generalization of
de Broglie-Bohm theory has produced so far is that it seems necessary to introduce a 
new element of reality, namely either a spacelike foliation of spacetime, or its dual 
concept, a timelike vector field on spacetime; see [\berndlETal], [\munchberndlETal].
        This is also the case in [\Kie], and in the model contemplated in [\Tumulka].
        The foliation would have to satisfy its own set of dynamical equations, 
which one can borrow from general relativity. 
        Where all this will lead us nobody knows, but hopefully the end-product will 
teach us something new about nature.
        
\smallskip
\noindent{\bf \quad Acknowledgement} 
I am indebted to Robert Rynasiewicz for inviting me to
contribute to the Bub-Festschrift, to Sheldon Goldstein 
for illuminating discussions and encouragement over many years, and
to Roderich Tumulka for comments on the ms.
Work supported by NSF Grant DMS-0406951. Any opinions,  
conclusions or recommendations expressed in this material are
those of the author and do not necessarily reflect the views of the NSF.

\biblio  

\bye